\documentclass[12pt]{iopart}
\usepackage[latin9]{inputenc}
\usepackage{amssymb}
\usepackage{graphicx}
\usepackage{esint}

\makeatletter
\usepackage{iopams}
\usepackage{setstack}

\makeatother

\begin{document}

\title[Electron-phonon relaxation and excited electron distribution ]{Electron-phonon relaxation and excited electron distribution in gallium nitride}

\author{ V.P.Zhukov$^{(1,2)}$, V.G.Tyuterev$^{(2,3,4)}$, P.M.Echenique $^{(2,5)}$ and E.V.Chulkov$^{(2,4,5)}$}

\address{$^{1}$ Institute of Solid State Chemistry, Urals Branch of the Russian
Academy of Sciences, Pervomayskaya st. 91, Yekaterinburg 620990, Russia}

\address{$^{2}$ Donostia International Physics Center (DIPC),P. Manuel de
Lardizabal 4, 20018 San Sebastian, Spain}

\address{$^{3}$ Tomsk State Pedagogical University, Kievskaya st. 60, Tomsk 634041,
Russia}

\address{$^{4}$ National Research Tomsk State University, Lenin St. 36,
Tomsk 634050, Russia}

\address{$^{5}$ Departamento de Fisica de Materiales, Facultad de Ciencias
Quimicas, UPV/EHU and CFM-MPC, Apartado 1072,20080 San Sebastian,
Spain}

\ead{tyuterev@gmail.com}
\begin{abstract}
We develop a theory of energy relaxation in semiconductors and insulators highly excited by the long-acting external irradiation. We derive the equation for the non-equilibrium distribution function of excited electrons. The solution for this function breaks up into the sum of two contributions. The low-energy contribution is concentrated in a narrow range near the bottom of the conduction band. It has the typical form of a Fermi distribution with an effective temperature and chemical potential. The effective temperature and chemical potential in this low-energy term are determined by the intensity of carriers' generation, the speed of electron-phonon relaxation, rates of inter-band recombination and electron capture on the defects. In addition, there is a substantial high-energy correction. This high-energy 'tail' covers largely the conduction band. The shape of the high-energy 'tail' strongly depends on the rate of electron-phonon relaxation but does not depend on the rates of recombination and trapping. We apply the theory to the calculation of a non-equilibrium distribution of electrons in irradiated GaN. Probabilities of optical excitations from the valence to conduction band and electron-phonon coupling probabilities in GaN were calculated by the density functional perturbation theory. Our calculation of both parts of distribution function in gallium nitride shows that when the speed of electron-phonon scattering is comparable with the rate of recombination and trapping then the contribution of the non-Fermi 'tail' is comparable with that of the low-energy Fermi-like component. So the high-energy contribution can affect essentially the charge transport in the irradiated and highly doped semiconductors.
\end{abstract}


\submitto{\SST}

\date{\today}
 \maketitle
\section{Introduction}

The study of relaxation processes in crystals is indispensable for the understanding of a number of physical phenomena induced by external sources at high excitation levels.  First of all this concerns to the experiments with the impact by electron beams and powerful lasers \cite{Otho98,Shah}. The knowledge of the energy and momentum relaxation is also indispensable for study of the high-field transport of charge carriers \cite{Brin98,Tisd10,Dur98}, including the case of electric breakdown \cite{Sun12,Bert09}. One should also mention such phenomena as laser ablation \cite{Wang11}, abrupt thermal impact \cite{Grin90,Gure02}, ultra-fast phase transitions \cite{Lind97a}, photocatalysis \cite{Zhuk14} and the solar energy conversion \cite{Nege06,Tea11}. 	In the investigation of the energy relaxation processes in $Si$ \cite{Otho98,Harb06}, $GaAs$ \cite{Tea11,Call00,Faso90,Supa96},
$GaP$ \cite{Coll13}, $InP$ \cite{Coll13,Hohe93} and $CdSe$ \cite{Prab95} the non-equilibrium distribution of electrons usually is simulated as a quasi-Fermi function which is localized close to the bottom the conduction band.

In our works \cite{Zhuk14,Zhuk12,Zhuk12a,Tyut15}  on the example of the compounds $ZnO$ and $TiO_2$ it was established that in a wide-bandgap material the evolution of energy relaxation has some peculiarities. It has been shown that the form of quasi-equilibrium distribution differs markedly from the case of narrow-bandgap compounds, and covers a wide range of energy within the conduction and valence bands over band edges at a distance comparable to the width of the band gap.

In the current article we will consider the quasi-equilibrium distribution in more general terms that allow us to incorporate into the theory \cite{Tyut15} the processes of radiative recombination and trapping of carriers. This makes it possible to specify the energy dependence of the distribution function either directly near the band edges as well as away from them. We apply the proposed theory to the case of gallium nitride in the wurtzite structure.

The paper is organized as follows. In Sec.\ref{Calc_met}, we present the analytical methods we use. First, we derive the equation for the quasi-equilibrium distribution function. Next, its formal solution is given in Subsec.\ref{Stat_solution}. On this basis in Subsec.\ref{eff_phonon} we develop an effective phonon model that allows an exact solution which is presented in Subsec.\ref{weak_deg}. Section \ref{techn} describes the technics of numerical calculation for $GaN$. The necessary material dependent quantities, namely, optical excitation probabilities and electron-phonon coupling characteristics are calculated \emph{ab initio} within the pseudopotential approach and density functional theory. Section \ref{applic} presents the application of Sec.\ref{Calc_met} and Sec.\ref{techn} to the non-equilibrium distribution of carriers in the conducting band of GaN. Finally, the results are discussed and summarized in Sec.\ref{concl}.

\section{Calculation methods\label{Calc_met}}

The kinetic equation for the time-dependent distribution function
$f(t,E)=G(E)n(t,E)$ of a non-equilibrium electron state created by
the powerful external source of high-energy irradiation was investigated
in Ref.{\cite{Tyut15}}. Here $G(E)=\sum_{c{\bf k}}\delta(E-E_{c{\bf k}})$
is the density of electron states in the conduction band, $n(t,E)$
is the occupation number of non-equilibrium electrons, the energy
$E$ is taken with respect of the bottom of conduction band (hereafter
'excess energy'). The excitation of electrons from the valence to
conduction band by external source is followed by the fast momentum randomization, within several femtoseconds \cite{Otho98}.
That is why the occupation number $n_{c\mathbf{k}}$ of electron's
band state $E_{c\mathbf{k}}$ can be regarded just as a function of
energy $n(t,E)$, the phonon occupation number is considered to be
energy-dependent as well: $N_{\sigma\mathbf{q}}=N(\epsilon)$.

We restrict ourselves to the case when the maximum excess energy of external excitation is located lower than the threshold of impact ionization $E_{imp}$. Numerically it roughly equals to the band-gap value above the bottom of the conduction band. Once the energy of an electron is lower than $E_{imp}$ the production of secondary electrons and holes is prohibited since it is impossible to simultaneously satisfy the energy and momentum conservation laws. Therefore inelastic electron-electron processes are absent in this range, and energy relaxation can only be associated with electron-phonon scattering.  For $GaN$ this threshold corresponds to the excess energy equal to 3.25 eV.

Retaining terms linear and quadratic by phonon energy which is a small
quantity relative to the band energy $E_{c{\bf k}}$ one comes to
the following equation for a non-equilibrium distribution $f(t,E)$
of electrons in the conduction band \cite{Tyut15}:

\begin{eqnarray}
 & \frac{df(t,E)}{dt}=\left[\frac{df(t,E)}{dt}\right]{}_{ext}-G(E)n(t,E)\gamma_{rec}(E)+\label{eq_N_stationar}\\
 & \frac{d}{dE}\left[\frac{dn(t,E)}{dE}\Gamma_{2}(E)+n(t,E)(1-n(t,E))\Gamma_{1}(E)\right]\nonumber
\end{eqnarray}
Hereafter
\begin{eqnarray}
 & \Gamma_{0}(E)=\int_{0}^{\epsilon_{m}}\Gamma(\epsilon,E)d\epsilon;\Gamma_{1}(E)=\int_{0}^{\epsilon_{m}}\epsilon\Gamma(\epsilon,E)d\epsilon\label{G0_G1}\\
 & \Gamma_{2}(E)=\int_{0}^{\epsilon_{m}}\epsilon^{2}(N(\epsilon)+1/2)\Gamma(\epsilon,E)d\epsilon\nonumber
\end{eqnarray}

A spectral function of electron-phonon interaction $\Gamma(\epsilon,E)$
is defined as:
\begin{eqnarray}
 & \Gamma(\epsilon,E)=\sum_{\sigma{\bf q}}\sum_{c{\bf k}c'{\bf k'}}\delta(E-E_{c{\bf k}})
 & P_{c{\bf k}c'{\bf k'}}^{\sigma{\bf q}}\delta_{{\bf k'\pm q-k}}\delta(E-E_{c'{\bf k'}})\delta(\epsilon-\epsilon_{\sigma{\bf q}})\label{G_spectral}
\end{eqnarray}
$P_{c{\bf k}c'{\bf k'}}^{\sigma{\bf q}}=\frac{2\pi}{\hbar}|\langle c{\bf k}|H_{\sigma{\bf q}}^{el-ph}|c'{\bf k'}\rangle|^{2}$
is the matrix element of electron-phonon interaction operator. $\epsilon_{\sigma{\bf q}}$
is the energy of a phonon of $\sigma^{th}$ branch with the wave vector
${\bf q}$. The term $[df(t,E)/dt]_{ext}$ describes the distribution
of excited carriers created by external sources of generation. The
expression $f(t,E)\gamma_{rec}(E)$ takes into consideration the flow
of electrons from the conduction band to the valence bands and to
the impurity levels. The range of integration in  Eqs.(\ref{G0_G1}) extends
from zero up to the maximal phonon energy $\epsilon_{m}$.
The function
\begin{equation}
\gamma_{e-ph}(E)=\Gamma_{0}(E)/G(E)\label{Probab_per_t}
\end{equation}
specifies the probability per unit time for an electron to leave a
given excess energy level $E$, so it is the electron-phonon relaxation
rate.

The well-known Eliashberg spectral function ${\alpha^{2}F}(\epsilon)$
of the electron-phonon interaction,  which relates to the Fermi-level in superconductors \cite{Eliashberg},
 can be generalized for an arbitrary excess energy level in semiconductor as
\begin{equation}
{\alpha^{2}F}(\epsilon,E)=\hbar\Gamma(\epsilon,E)/(2\pi G(E))\label{Eliash1}
\end{equation}
This function, in its turn, determines in semiconductor the
electron-phonon coupling constant $\lambda(E)$, relating to a given energy $E$ :
\begin{equation}
\lambda(E)=2\int_{0}^{\epsilon_{m}}d\epsilon{\alpha^{2}F}(\epsilon,E)/\epsilon\label{Coupl_const}
\end{equation}

\subsection{Stationary solution for the distribution function\label{Stat_solution}}

If the action of external excitation continues for a time long enough
then the quasi-stationary distribution become established. Then a
search of a quasi-stationary distribution function $df(t,E)/dt=0$
reduces itself to the solution of the equation for the time-independent
occupation number $n(E)$. It is convenient to seek for a stationary solution
of (\ref{eq_N_stationar}) in the form $n(E)=n_{0}(E)+n_{1}(E)$.
Here $n_{0}(E)$ is the solution for the stationary equation (\ref{eq_N_stationar})
but without sources and sinks. One can check by a direct substitution
that there exists a partial solution for $n_{0}(E)$ in the following
form
\begin{equation}
n_{0}(E)=(e^{R(E)}+1)^{-1}\label{N0_general}
\end{equation}
where
\begin{equation}
R(E)=\int_{0}^{E}\Gamma_{1}(E')/\Gamma_{2}(E')dE'+C\label{R_general}
\end{equation}
Here $C$ is an integration constant.

Under conditions of low intensity of external exposure one has to
regard $n_{0}(E)$ as a small quantity. In this case keeping the leading
terms we come from Eq.(\ref{eq_N_stationar}) to the linearized equation
with a stationary solution
\begin{equation}
n_{0}(E)=e^{-R(E)}\label{n0_quasiBoltz}
\end{equation}
Both distributions (\ref{N0_general}) and (\ref{n0_quasiBoltz})
should be localized in the energy region where the electron-phonon
relaxation is inefficient, i.e. at $E\leq\epsilon_{m}$, where $\epsilon_{m}$
is the maximal energy of the phonon spectrum. An equation for the
correction $n_{1}(E)$ comes out by substitution of $n_{0}(E)$ into
Eq.(\ref{eq_N_stationar}). We restrict ourselves by the examination
of excitation regimes when $n_{1}(E)$ can be treated as a small and
smooth quantity. Retaining the leading terms the equation for the
correction term in the stationary regime acquires the following form:
\begin{equation}
\frac{d}{dE}\{[1-2n_{0}(E)]n_{1}(E)\Gamma_{1}(E)\}=
 n_{0}(E)G(E)\gamma_{rec}(E)-[df(t,E)/dt]_{ext} \label{eq_for_n1}
\end{equation}
The solution of
Eq.(\ref{eq_for_n1}) which satisfies the condition $n_{1}(E)$= 0
at $E\geq E_{max}$ can be written as
\begin{eqnarray}
 & n_{1}(E)=\{[1-2n_{0}(E)]\Gamma_{1}(E)\}^{-1}\times\label{Solut_n1_gener-1}\\
 &\times \int_{E}^{E_{max}}\{[df(t,E')/dt]_{ext}-n_{0}(E')G(E')\gamma_{rec}(E')\}dE'\nonumber
\end{eqnarray}
Here $E_{max}$ is the maximal level of electrons excitation.

In a stationary regime the total number of electrons generated per
unit time by the external source should be equal to the number of
electrons that leave the conduction band during the same time due
to the inter-band recombination and trapping by the impurity levels:
\begin{eqnarray}
 & \int_{0}^{E_{max}}[df(t,E)/dt]_{ext}dE=
 & \int_{0}^{E_{max}}n_{0}(E)G(E)\gamma_{rec}(E)dE\label{Diverg_gener-1}
\end{eqnarray}
Note, that $\Gamma_{1}(E)\sim E$ as $E\rightarrow0$ . Hence Eq.(\ref{Diverg_gener-1})
provides the proper finite limit of $n_{1}(E)$ at the bottom of the
conduction band.
This equation defines implicitly the integration constant $C$ in Eq.({\ref{R_general})

As $n_{0}(E)$ is mostly localized in a small region of order of phonon spectrum
width $E<\epsilon_{m}$ above the conduction band bottom then one can neglect
the value $n_{0}(E)\ll1$ at $E>\epsilon_{m}$.

The non-equilibrium occupation number $n_1(E)$
in the high energy region can be derived using Eq.(\ref{Diverg_gener-1})
which determines the integration constant $C$ in the solution
(\ref{N0_general}) or (\ref{n0_quasiBoltz}) of $n_{0}(E)$.
\begin{equation}
n_{1}(E)\approx\Gamma_{1}^{-1}(E)\int_{E}^{E_{max}}[df(t,E')/dt]_{ext}dE'\label{Quasi_equillibrium_part}
\end{equation}
The expression similar to Eq.(\ref{Quasi_equillibrium_part}) for
the high energy 'tail' of the distribution function $f_{1}(E)=n_{1}(E)G(E)$
was formerly discussed in Refs.\cite{Zhuk12,Zhuk12a,Zhuk14} where
the electron-hole recombination has not been formally taken into consideration.
One can see now that the electron-hole recombination actually does
not affect the shape of the high-energy 'tail' of distribution.

\subsection{The 'effective phonon' model\label{eff_phonon}}

Let us define an averaged energy loss per one electron in
the process of a single transition act from an excess level $E$ as
\begin{equation}
\epsilon_{av}(E)=\Gamma_{1}(E)/\Gamma_{0}(E)\label{aver_phonon__energy-1}
\end{equation}
In calculations of Refs.\cite{Zhuk12,Zhuk12a} it was shown that
for realistic band structures in $ZnO$ and $TiO_{2}$ this quantity
 manifests a weak energy dependence. Below we will show that this is also valid
 for $GaN$. Hence we neglect this dependence and introduce the Einstein-like lattice vibrational model considering
the electron scattering by a single 'effective' phonon with energy
$\epsilon_{0}$, and identify $\epsilon_{0}$ with $\epsilon_{av}$.
In this model we assume that
\begin{equation}
\Gamma(\epsilon,E)=P_{av}(E)G^{2}(E)\delta(\epsilon-\epsilon_{0})\label{Effective-phonon}
\end{equation}

Here we introduce the averaged electron-phonon coupling factor
\begin{equation}
P_{av}(E)=\gamma_{e-ph}(E)/G(E)=\Gamma_{0}(E)/G^{2}(E)\label{Aver_P-1}
\end{equation}

Then the definition (\ref{Probab_per_t}) implies that
\begin{equation}
\gamma_{e-ph}(E)=G(E)P_{av}(E)\label{Prob_per_t_efPhon}
\end{equation}

Expressions for $\Gamma_{0}$, $\Gamma_{1}$,$\Gamma_{2}$ follow
from definitions (\ref{G0_G1}):
\begin{eqnarray}
 & \Gamma_{0}(E)=P_{av}(E)G^{2}(E);\nonumber \\
 & \Gamma_{1}(E)=\epsilon_{0}P_{av}(E)G^{2}(E);\label{Gamm_EffPhon}\\
 & \Gamma_{2}(E)=\epsilon_{0}^{2}(N(\epsilon_{0})+1/2)P_{av}(E)G^{2}(E)\nonumber
\end{eqnarray}

Taking these approximations into account we find also the expression
for $R(E)$ as defined by the Eq.(\ref{R_general})
\begin{eqnarray}
R(E)=E/[\epsilon_{0}(N(\epsilon_{0})+1/2)]+C\label{R_effPhon}
\end{eqnarray}
Then Eq.(\ref{N0_general}) for $n_{0}$ acquires the form of
quasi-Fermi distribution
\begin{eqnarray}
n_{0}(E)=(\exp(\frac{E-\mu}{k_{B}T_{eff}})+1)^{-1}\label{N0_Fermy}
\end{eqnarray}
where effective temperature of the excited electrons near the bottom
of the conduction band is $T_{eff}=\epsilon_{0}(N(\epsilon_{0})+1/2)/k_{B}$.
The effective chemical potential $\mu=-k_{B}T_{eff}C$ is defined implicitly by equation (\ref{Diverg_gener-1}).

\subsection{The case of weak degeneracy\label{weak_deg}}

We discuss hereafter the case of week degeneracy, when $n_{0}(E)\ll1$.
This is a realistic case for the exposure by the sunlight or a mercury
lamp\cite{Loud83}, when the typical number of photons per unit cell of a crystal
 does not exceed $10^{-5}$. In this case Eq.(\ref{n0_quasiBoltz}) reduces to the quasi-Boltzmann distribution:
\begin{eqnarray}
n_{0}(E)=A\exp(-E/[\epsilon_{0}(N(\epsilon_{0})+1/2)])\label{n0-Boltzmman-efPh}
\end{eqnarray}
For further simplification  let us introduce the separable approximation for the emission term
$[df(t,E)/dt]_{ext}$ in Eq.(\ref{Quasi_equillibrium_part})
\begin{eqnarray}
[df(t,E)/dt]_{ext}=S_{0}(t)\overline{S}(E)\label{Nakachka}
\end{eqnarray}
Here the function $S_{0}(t)$, in general time-dependent, is defined
to be equal to the total number of excited electrons per unit of time,
that depends on the intensity of exposure. In the quasi-stationary
regime $S_{0}(t)$ changes slowly, $dS_{0}/dt\ll S_{0}(t)/\tau_{e-ph}$, and
can be considered as a constant $S_{0}(t)\thickapprox S_{0}$.
The function $\overline{S}(E)$ describes the distribution of the
excited electrons over the energy scale of the conduction band. It
has to be normalized to unity: $\int\overline{S}(E)dE=1$. Employing
this definition and effective phonon approximation one can write the
expression for the 'tail' distribution as
\begin{equation}
n_{1}(E)\approx\frac{S_{0}}{\epsilon_{0}\gamma_{e-ph}(E)G(E)}\int_{E}^{E_{max}}\overline{S}(E')dE'\label{n1_S0}
\end{equation}

The first-principle calculations demonstrate that in $GaN$ near the
bottom of the conduction band $G(E)$ has almost linear energy dependence.
We approximate this dependence as $G(E)=G_{0}E$.

In general, $N(\epsilon_{0})$ is also a non-equilibrium distribution of phonons, and it would be necessary to investigate its evolution in conjunction with the evolution of the electron distribution. However, if the duration of action of external excitation exceeds the characteristic relaxation time of the phonon subsystem, it can be regarded as Boltzmann function.

In the low-temperature
limit when $N(\epsilon_{0})\ll1$ we can neglect a phonon occupation number. We
 also neglect the energy dependence of the recombination and trapping
rate $\gamma_{rec}(E)$ in the small energy range $\epsilon_{m}$
near the bottom of the conduction band, hence $\gamma_{rec}(E)=\gamma_{rec}$. In the high excitation
regime the value $E_{max}\gg\epsilon_{m}$ and in Eq. (\ref{Diverg_gener-1})
it can be replaced by infinity. This helps us, employing Eq.(\ref{Diverg_gener-1}),
to find the normalization coefficient $A$ in expression (\ref{n0-Boltzmman-efPh})
for $n_{0}$:
\begin{equation}
A=4S_{0}/(\gamma_{rec}G_{0}\epsilon_{0}^{2})\label{Const_A}
\end{equation}
Now we see that the $n_{0}(E)$ depends on the intensity of exposure,
via the function $S_{0}$, as well as on the rate $\gamma_{rec}$ of
trapping and recombination processes.

Hence the total electron distribution function
\begin{equation}
f(E)=f_{0}(E)+f_{1}(E)=G(E)(n_{0}(E)+n_{1}(E))\label{f0pluf1}
\end{equation}
is written as
\begin{eqnarray}
&f(E)=S_{0}[\left\{ 4\gamma _{rec}^{-1}E/\epsilon _{0}-\gamma_{e-ph}^{-1}\left( E\right) \right\} \times\\\nonumber
&\times \exp \left( -2E/\epsilon_{0}\right) +\gamma _{e-ph}^{-1}\left( E\right)\int_{E}^{E_{max}}\overline{S
}\left( E\right) dE]/\epsilon _{0}\label{f_E-full}
\end{eqnarray}

\section{Technical details\label{techn}}

We apply the described approach to the gallium nitride compound in the wurtzite
structure. Our calculations have been performed by employing the \emph{ab initio}
 pseudo-potential computer code Quantum Espresso (QE) based
on the density functional theory for the electron band structure and
density-functional perturbation theory for the phonon band states
\cite{Baro01}. The pseudo-wave functions were expanded in plane
waves with energy cutoff 820 eV.
In the calculations of the electron states we employed
a set of 50 wave vectors per irreducible part of the Brillouin zone
(IPBZ).  The calculations
of the phonon states and characteristics of the electron-phonon coupling
were performed for a set of 12 wave vector in the IPBZ. Gallium
 norm-conserving atomic pseudo-potential was calculated using the  Bachelet-Hamann-Schlüter method \cite{Bach82} with Perdew-Zunger  exchange-correlation potential \cite{Perd81}. The pseudo-potential for nitrogen was
constructed within the approach of Von Barth and Car \cite{DalCors93}. We will show later that such
a way of calculations provide very good results both for electron
and phonon band structures.

We compute the $\overline{S}(E)$ distribution function also basing
on the band structure calculations by means of the QE computer code.
Namely, if the energy of the quantum of optical excitation is $E_{exc}$
then for the excess energy $E$ we sum the probabilities of all
direct excitations from the electronic states at the energy $E-E_{exc}$
to the states at the energy $E$. Hence, the non-normalized $\overline{S}(E)$
function is
\begin{equation}
\overline{S}(E)=\sum_{{\bf k}nn'}\delta(E-E_{n{\bf k}})T(n{\bf k},n'{\bf k})\delta(E-E_{exc}-E_{n'{\bf k}})\label{eq37}
\end{equation}
where $T(n{\bf k},n'{\bf k})$ is the probability of the transition
between the states $|n{\bf k}\rangle$ and $|n'{\bf k}\rangle$. For
the calculations of the $T(n{\bf k},n'{\bf k})$ transition probabilities
we apply dipole approximation. The details of such an approach can
be found in Ref.\cite{Bena}.

\section{Results and discussions\label{applic}}

The calculated dispersion curves and the corresponding total density of electron states
are given for $GaN$ in Fig.\ref{Fig1}.
The direct band gap in $\Gamma$-point is equal to 3.25 eV, in good agreement
with experimental value of 3.4 eV (Ref.\cite{Stre00}). The top of the valence band is formed by $2p$-like states of nitrogen, the bottom of the conduction band corresponds to the $4s$-like states of $Ga$.
 Also the energy differences
between the conduction band states at symmetry points of the Brilloiune
zone agree with the results of previous calculations
\cite{Bulu00}. E.g., our calculated $\Gamma_{1}-K_{2}$, $\Gamma_{1}-L_{1,3}$,
$\Gamma_{1}-M_{1}$ differences are equal to 1.8, 1.55, 2.06 eV, whereas
Bulutay et al.\cite{Bulu00} obtain 1.6, 1.75,
1.87 eV, respectively. The conduction band at $\Gamma_{1}$ displays the non-parabolic
dispersion, so the density of states versus the excess electron energy
deviates from the free-electron-like law $\sim\surd\overline{E}$
practically immediately above the bottom of the conduction band.
The non-parabolicity of the conduction band states has been revealed
experimentally in Ref.\cite{Syed03}. Up to the excess energy of
about 1.25 eV, that is up to the energy in the central $\Gamma_{1}$
valley lying between the side $\L_{1,3}$ and $M_{1}$ states, the
total DOS displays almost linear dependence (see for inset in Fig.\ref{Fig1}),
so near the bottom of the conduction band we approximate
it with $G(E)=G_{0}E$ where $G_{0}$=0.025$eV^{-2}$.
\begin{figure}[h]
\centering\includegraphics[width=0.48\textwidth]{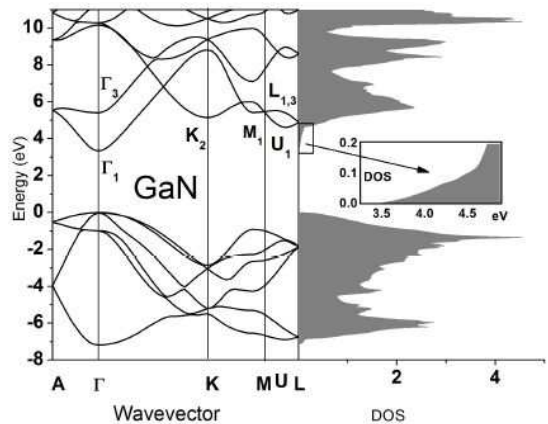}\caption{Electron dispersion curves and density of states for GaN. The inset displays DOS at the
bottom of a conduction band}
\label{Fig1}
\end{figure}

In Fig.\ref{Fig2} our calculated phonon dispersion curves are compared
with the experimental data of Ref.\cite{Ruf01}. Similarly to the
theoretical results of this paper, our computed frequencies are systematically about 5 \% higher than
the experimental ones. We demonstrate
in this figure the calculated data scaled with the factor 0.95.
 In Fig.\ref{Fig2} we also show the corresponding density of phonon states. In the following calculations the highest energy of the phonon spectrum is taken to be equal to 785 $cm^{-1}$ = 0.097 eV.
\begin{figure}[h]
\centering\includegraphics[width=0.48\textwidth]{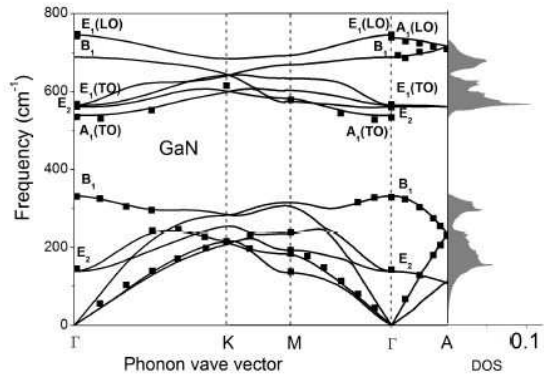}\caption{Phonon dispersion curves and DOS for GaN. Experimental data are taken
from Ref.\cite{Ruf01}}
\label{Fig2}
\end{figure}

Hereafter we discuss the electron-phonon relaxation at excess energies which do not exceed the impact ionization level 3.25 eV in $GaN$. The inter-valley electron-phonon scattering events happen at
higher energies and are not a subject of the current study.

In Fig.\ref{Fig3} we show the Eliashberg function calculated for
three different values of excess energy in interval from 0.03 to 1.83 eV. We
see that with increasing excess energy the electron-phonon scattering
becomes stronger. For the states near the bottom of the conduction band the scattering
of excess electrons energy occurs mainly via emission of optical
phonons, both transversal ($TO$) and longitudinal ($LO$) branches, being
in correspondence with the common opinion on the main role of such
phonons in the processes of electron energy relaxation \cite{Lee97}.
We see, however, that with rising electron excess energy the contribution
of acoustic branches essentially increases. In contrast to Ref.\cite{Stan01} where
the dominating scattering via emission of the $LO$ phonons was supposed to occur, we find
that at any value of the excess energy the contribution of the $TO$-branches
is almost equal to that of longitudinal ones.
\begin{figure}[h]
\centering\includegraphics[width=0.45\textwidth]{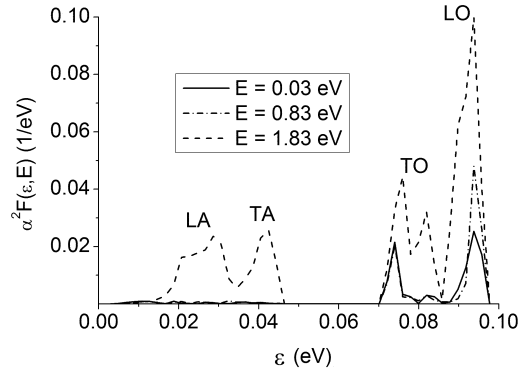} \caption{Generalized Eliashberg function $\alpha^{2}F(\varepsilon,E)$
for three different values of the electron excess energy }
\label{Fig3}
\end{figure}

In Fig.\ref{Fig4} we demonstrate the energy dependence of the $\lambda$,
$\gamma_{e-ph}$, $\epsilon_{av}$ values. The dependencies of $\lambda$
and $\gamma_{e-ph}$also demonstrate the strengthening of electron-phonon
scattering with the rise of the excess energy. However, the effective
phonon energy $\epsilon_{av}$ only slightly changes, so in the estimation
of the relaxation rate and distribution curves we use the energy-average
value $\epsilon_{0}$ = 0.085 eV. We see in the Eliashberg
curves that the $\epsilon_{0}$ value lies just
 between the bands of the $LO$ and $TO$ branches which
at any value of the excess energy have almost equal contributions
to the electron relaxation rate.
\begin{figure}[h]
\centering\includegraphics[width=0.48\textwidth]{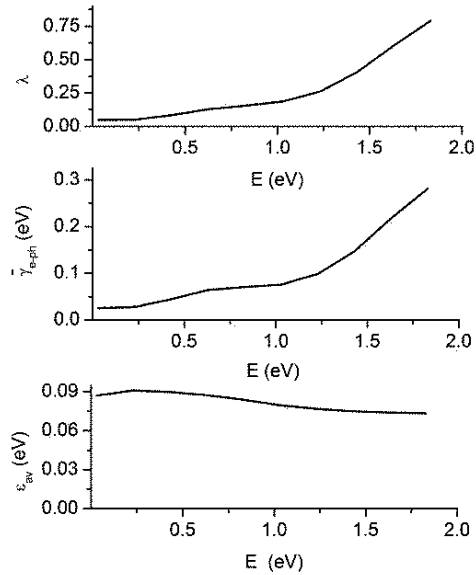}
\caption{Energy dependence of the constant of electron-phonon scattering $\lambda$,
the rate of the scattering $\gamma_{e-ph}$ and the averaged value
of the emitted phonon energy $\epsilon_{av}$.}
\label{Fig4}
\end{figure}

According to Eq.(\ref{Prob_per_t_efPhon}) the change of average lifetime
of a single electron
$\tau_{e-ph}(E)=\gamma_{e-ph}^{-1}(E)$
with energy is determined by the changes of the density of states
$G(E)$ and the averaged probability of phonon emission $P_{av}(E)$.
In Fig.\ref{Fig5} we show $\tau_{e-ph}(E)$,  $G(E)$ and
$P_{av}(E)$ versus the excess energy.

We see very rapid change of $P_{av}(E)$ at the energy
less than $\sim$ 0.25 eV. But at a higher energy  $P_{av}(E)$
rather slowly decreases with energy while $G(E)$ rapidly increases
 that leads to the reduction of $\tau$. In other
words, the reduction of the electron relaxation time is determined
by the expansion of the space of electronic states available for the
electron that looses its energy via phonon emission. Note that the
changes of $P_{av}(E)$ in Fig.\ref{Fig5} are very similar to that
observed in our previous study of the electron dynamics\cite{Zhuk12a}.
\begin{figure}[h]
\centering\includegraphics[width=0.45\textwidth]{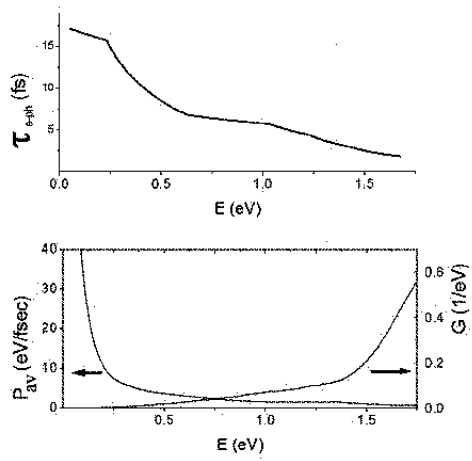}\caption{ Energy dependencies of the electron relaxation time $\tau_{e-ph}(E)$,
averaged rate of the single phonon emission $P_{av}(E)$ and density
of the electron states $G(E)$. }
\label{Fig5}
\end{figure}

The available literature data on the electron relaxation time in $GaN$
are contradictory. In Refs.\cite{Stan01, Tsen97} the electron-phonon
relaxation time has been evaluated using Fröhlich
theory of electron-phonon scattering which employs
the experimental values of dielectric constants $\varepsilon(\infty),\varepsilon(0)$
and phonon $\omega_{LO},\omega_{TO}$ frequencies.
This estimation of relaxation time, 10 fs, corresponds well
to our energy-averaged data. However, the
energy dependence of the relaxation time has not been investigated.

The energy dependence of a relaxation time was investigated
in  Refs.\cite{Bulu00,Dasg13}. Bulutay et al \cite{Bulu00},
also relying on the Fröhlich's electron-phonon
matrix elements, performed the calculations
 of phonon-assisted scattering rate in $GaN$ taking into account the coupling
 of excited electrons with long-wave $LO$ and $TO$ phonon modes. According
to these calculations, the relaxation rate increases with the excess
energy, although this growth is not as rapid as in our first-principle
calculations. Dasgupta et al.\cite{Dasg13}  in their
experimental work
came to the conclusion that the relaxation rate decreases with excess
energy. However, their estimation of the relaxation rate and time
was not direct since they were based on the measurements of the inelastic
mean free path (IMFP) of electrons in a hot electron transistor with
$GaN$ as the base layer. The increase of the IMFP with the excess energy
can be associated both with the increase of the electron velocity
and the reduction of the electron-defect scattering. The discrepancy between our ab initio calculated
data and experimental data of Ref.\cite{Dasg13} deserves a special
study.

Let us note in this connection that Fröhlich's interaction is limited to the interaction with the long-wavelength optical phonon. So the phenomenological electron-phonon Fröhlich's Hamiltonian involves only comparatively small values of the wave vector. That could be enough in the case of a weak excitation because the scattering events are concentrated at the bottom of the conduction band. In our study of large deviation from equilibrium, we should not be limited to a single long-wavelength optical phonon, instead we should take into account all phonon states, including all nine optical branches over the entire Brillouin zone.

  In Fig. \ref{Fig6} the results of our calculations for three
distribution curves, $\overline{S}(E)$ , $f_{0}(E)$ and $f_{1}(E)$
are shown. Our aim is to compare values of $f_{0}(E)$ and $f_{1}(E)$,
so we omit the factor $S_{0}(t)/\epsilon_{0}$. Two parameters in
our evaluations are determined by the experimental
conditions, namely the excitation energy $E_{exc}$ and the rate of electron-hole
recombination and trapping $\gamma_{rec}$. We analyse
the variation of the distribution functions with respect to $E_{exc}$
by calculating the $\overline{S}(E)$ and $f_{1}(E)$ functions for
three values of $E_{exc}$: 3.48, 4.08 and 5.08 eV, that correspond
to the lower edge, the center and the top of our energy
interval of interest. The value of $\gamma_{rec}$ is a 'technological'
parameter that strongly depends on the concentration of impurities
and intrinsic defects in a crystal. We study its influence on the
distribution functions by varying it from 0.04 eV, the value typical
for the rate of electron-phonon scattering, up to the values of
10 and 100 times less, that is going to the pico-second times of recombination
and trapping processes.

We see that the function $\overline{S}(E)$ has a simple gaussian shape
with a peak energy increasing almost linearly with
a rise of $E$, $E_{exc}$. Respectively, with increasing $E_{exc}$
 the upper boundary of the distribution $f_{1}(E)$ shifts to a higher energy, but
the values of the $f_{1}(E)$ function do not change markedly, keeping
at the level about 40 $eV^{-1}$. The functions $\overline{S}(E)$ and $f_{1}(E)$
of $GaN$ appear to be similar to those calculated before for $ZnO$
\cite{Zhuk12a}, in spite of a different way of calculating,
 based on the LMTO approach.
We also see that when $\gamma_{rec}$ is taken to be close to the rate
of electron-phonon scattering $\gamma_{e-ph}$, then the values of $f_{1}(E)$
and $f_{0}(E)$ appear to be comparable. When $\gamma_{rec}$ is 10
times smaller than $\gamma_{e-ph}(E)$, then $f_{1}(E)$ is also
smaller than  $f_{0}(E)$, but still not negligible. When
$\gamma_{rec}$ is of two orders lower than $\gamma_{e-ph}(E)$,
then the value of $f_{1}(E)$ becomes relatively small, it does not exceeds
1.5\% of $f_{0}(E)$.
\begin{figure}[h]
\centering\includegraphics[width=0.45\textwidth]{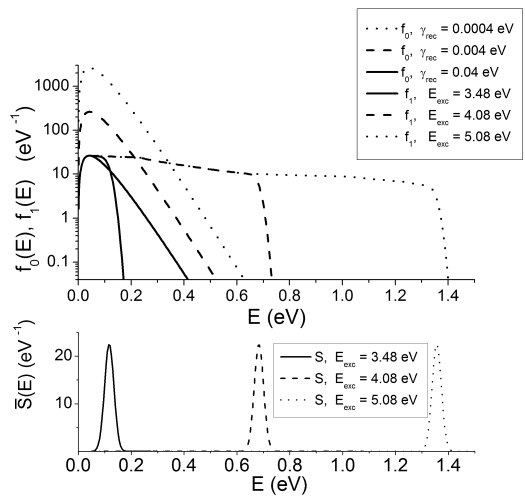} \caption{The energy dependencies of the $f_{0}$, $f_{1}$, $\overline{S}$
distribution functions }
\label{Fig6}
\end{figure}

\section{Conclusions\label{concl}}
On the basis on the kinetic theory we derived the equation for the quasi-equilibrium distribution function of electrons in semiconductors excited by strong long-lasting external irradiation. The leading role in the formation of the carriers' distribution is associated with inelastic scattering of electrons by phonons.  The solution essentially describe a stable flow of electrons downward in an energy scale. In the case of wide-bandgap material we can allocate two parts. One of them $f_{0}(E)$ describes the accumulation of electrons above the bottom of the conduction band and has a conventional Fermi-like shape with a width of order of phonon spectrum. Its size depends both on the level of excitation and on the rate of 'leakage' of electrons through the process of electron-hole recombination and electron trapping by the defects. The effective temperature and chemical potential generally depend on an occupation number of excited optical phonon.

When the rate of recombination of electrons and holes and capture electrons on defects is much smaller than the characteristic time of electron-phonon relaxation then a kind of 'bottleneck' arises in that flow.
The high-energy 'tail' $f_{1}(E)$
appears as a result of a balance between the supply from an external source and leakage.  This second contribution is much smaller in size but extends much further over the conduction band.The form of the 'tail' depends on the band structure and on the rate of electron-phonon energy relaxation. Interestingly, its shape and scale do not depend on the rate of recombination and trapping. The depth of their penetration into the conduction band depends on the rate of excitation of electrons from the valence band to the conduction one and can be comparable with the magnitude of the band gap $E_g$ above the bottom of the conduction band.

Based on the theory we have carried out the first-principle calculations for the characteristics of electron-phonon scattering and the distribution functions for $GaN$ in the wurtzite structure. A good quality of the calculated electron and phonon band structures is confirmed by the comparison with available experimental and theoretical data. By calculating the spectral distribution of the electron-phonon scattering (Eliashberg function) we have revealed that the contribution of the LO-phonon branch to the electron relaxation rate is comparable with that of the TO-phonon branch; the contribution of the acoustic branch, although markedly smaller, is not negligible. We also find that the electron-phonon relaxation time decreases with the rise of the electron excess energy, which is determined by the increase of the electron density of states.

Calculating the shape of electron distributions we find that when the rate of electron-phonon scattering is comparable with the rate of recombination and electron trapping by lattice defects then the values of the low-energy component are comparable with those of the high-energy 'tail' of distribution. However, when the rate of recombination decreases then the value of the low-energy term becomes bigger. When the rate of electron recombination/trapping is two orders smaller, i.e. these processes occur in the picosecond energy range, the tail distribution does not exceed 1.5 \% of the low-energy distribution.

Considering $GaN$ as a typical example we conclude that the high-energy correction   has to be taken into account in the study of the behavior of highly excited charge carrier in a wide-bandgap semiconductor compounds owing a high rate of electron-phonon scattering. The length of the high-energy 'tail'  may cover a substantial part of the conduction band. Its contribution to the total concentration of non-equilibrium carriers may be comparable to that of the conventional Fermi part and thus can have a significant influence on the transport properties of the irradiated crystal.

\ack
 The work was supported by RFBR, research project No.15-02-00293  and the Tomsk State
University Competitiveness Improvement Program. V.G.T. acknowledges the support of Ministry of Education and Science of Russian Federation base part 101.
 The calculations have been performed using the URAN cluster of the Institute of mathematics
and mechanics in the Urals branch of the Russian Academy of Sciences
and SKIF CYBERIA cluster of Tomsk State University.



\end{document}